\documentclass[english,letterpaper,twocolumn,american,prl,superscriptaddress,showpacs]{revtex4}
\usepackage[T1]{fontenc}
\usepackage[latin9]{inputenc}
\usepackage{graphicx}

\makeatletter

\newcommand{\noun}[1]{\textsc{#1}}

\@ifundefined{textcolor}{}
{%
 \definecolor{BLACK}{gray}{0}
 \definecolor{WHITE}{gray}{1}
 \definecolor{RED}{rgb}{1,0,0}
 \definecolor{GREEN}{rgb}{0,1,0}
 \definecolor{BLUE}{rgb}{0,0,1}
 \definecolor{CYAN}{cmyk}{1,0,0,0}
 \definecolor{MAGENTA}{cmyk}{0,1,0,0}
 \definecolor{YELLOW}{cmyk}{0,0,1,0}
 }

\makeatother

\usepackage{babel}

\begin{document}

\title{Conductance of Atomic-Sized Lead Contacts in an Electrochemical Environment}

\author{F.-Q. Xie}

\affiliation{Institut für Angewandte Physik and DFG Center for Functional Nanostructures,
Karlsruhe Institute of Technology, 76131 Karlsruhe, Germany}

\author{F. Hüser}

\affiliation{Institut für Theoretische Festkörperphysik and DFG Center for Functional
Nanostructures, Karlsruhe Institute of Technology, 76131 Karlsruhe,
Germany}

\author{F. Pauly}

\affiliation{Institut für Theoretische Festkörperphysik and DFG Center for Functional
Nanostructures, Karlsruhe Institute of Technology, 76131 Karlsruhe,
Germany}

\email{fabian.pauly@kit.edu}

\author{Ch. Obermair}

\affiliation{Institut für Angewandte Physik and DFG Center for Functional Nanostructures,
Karlsruhe Institute of Technology, 76131 Karlsruhe, Germany}

\author{G. Schön}

\affiliation{Institut für Theoretische Festkörperphysik and DFG Center for Functional
Nanostructures, Karlsruhe Institute of Technology, 76131 Karlsruhe,
Germany}

\affiliation{Institut für Nanotechnologie, Karlsruhe Institute of Technology,
76344 Eggenstein-Leopoldshafen, Germany}

\author{Th. Schimmel}

\affiliation{Institut für Angewandte Physik and DFG Center for Functional Nanostructures,
Karlsruhe Institute of Technology, 76131 Karlsruhe, Germany}

\affiliation{Institut für Nanotechnologie, Karlsruhe Institute of Technology,
76344 Eggenstein-Leopoldshafen, Germany}

\email{thomas.schimmel@kit.edu}

\date{\today}
\begin{abstract}
Atomic-sized lead (Pb) contacts are deposited and dissolved in an
electrochemical environment, and their transport properties are measured.
Due to the electrochemical fabrication process, we obtain mechanically
unstrained contacts and conductance histograms with sharply resolved,
individual peaks. Charge transport calculations based on density functional
theory (DFT) for various ideal Pb contact geometries are in good agreement
with the experimental results. Depending on the atomic configuration,
single-atom-wide contacts of one and the same metal yield very different
conductance values.
\end{abstract}


\maketitle
Charge transport through nanostructures presently constitutes a highly
active area of research, also due to its relevance for the further
miniaturization of electronic devices \cite{Agrait:PhysRep2003}.
Regarding atomic-sized metallic contacts, different experimental techniques
have been developed for their fabrication, e.g.\ mechanically controllable
break junctions (MCBJs) \cite{Muller:PhysicaC1992}, modified scanning
tunneling microscopes (STMs) \cite{Gimzewski:PRB1987}, electromigration
\cite{Park:ApplPhysLett1999}, and electrochemical methods \cite{Li:Nanotechnology1999,Li:APL2002}.
While in MCBJ experiments and STM-based setups atomic-sized contacts
are formed by plastic deformation of the contact area, the advantage
of the electrochemical deposition method is the absence of mechanical
strain during contact formation, resulting in corresponding differences
in conductance histograms. 

Conductance histograms of $s$-valent metals, such as gold, silver,
or copper, have been analyzed extensively \cite{Hansen:PRB1997,Ludoph:PRB2000CondFluct,Agrait:PhysRep2003},
showing a preference for multiples of $G_{0}=2e^{2}/h$, with the
peak at $1G_{0}$ being a very robust feature. As it is known that
the chemical valence plays an important role for electron transport
\cite{Scheer:Nature1998,Pauly:PRB2006}, conductance histograms for
multivalent metals may consequently show a very different shape \cite{Ludoph:PRB2000CondFluct,Agrait:PhysRep2003}.

In this work, Pb will serve as a model system for multivalent metals.
For Pb a broad peak between $1G_{0}$ and $3G_{0}$ was observed in
conductance histograms based on the MCBJ technique \cite{Yanson:PhD2001,Makk:PRB2008}.
During stretching of the junctions, the plateau regions of the conductance
traces commonly exhibit a negative slope as a result of the mechanical
strain \cite{Cuevas:PRL1998,Scheer:Nature1998}. With electrochemical
techniques we have recently fabricated atomic-scale silver contacts
with a high thermal and mechanical stability \cite{Obermair:Nato2004},
and exploited this for constructing gate-controlled atomic switches
\cite{Obermair:PRL2004}. In this work, we study charge transport
through atomic-sized Pb contacts in experiment and theory. The conductance
histogram is compared to the literature and theoretical results for
ideal contact geometries. In the DFT-based calculations we explore
both {}``single-atom'' and {}``dimer'' geometries with a single
atom or a chain of two atoms in the narrowest part of the junction,
respectively. For these configurations we consider orientations of
the semi-infinite electrodes along the three main crystallographic
directions, namely $\left\langle 100\right\rangle $, $\left\langle 110\right\rangle $,
and $\left\langle 111\right\rangle $.

\begin{figure}
\begin{centering}
\includegraphics[width=0.8\columnwidth]{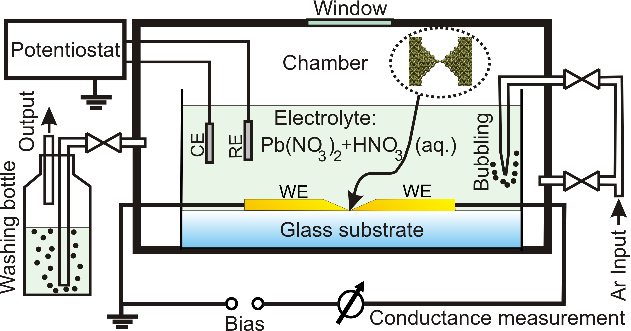}
\par\end{centering}

\caption{\label{fig:setup}(Color online) Schematic diagram of the experimental
setup. Pb is deposited and dissolved electrochemically in the narrow
gap between two gold working electrodes (WEs) on a glass substrate,
and the conductance of the contact is measured simultaneously. The
potentials of the WEs with respect to the lead reference electrode
(RE) and the lead counter electrode (CE) are set by a potentiostat.}

\end{figure}

Our experimental setup is illustrated schematically in Fig.~\ref{fig:setup}.%
 Since lead oxidizes under ambient conditions, the electrochemical
cell is shielded in an inert gas chamber. Before starting with the
electrochemical deposition, the chamber is thoroughly streamed with
Ar. In this way, a practically oxygen-free environment is created
for the growth of the atomic-sized Pb contacts. For their fabrication,
two working electrodes (WEs) made of gold and separated by a gap of
approximately 50 nm are prepared. They are covered with an insulating
polymer coating except for the immediate contact area to minimize
ionic conduction. Two lead wires (0.5 mm diameter, 99.998\% purity)
are used for the quasi-reference electrode (RE) and the counter electrode
(CE). The potentials of the WEs with respect to the RE and the CE
are set by a computer-controlled potentiostat. The electrolyte consists
of 1mM Pb(NO$_{3}$)$_{2}+0.1$ M HNO$_{3}$ in bi-distilled water.
For conductance measurements an additional voltage of -12.9 mV is
applied between the two WEs. To exclude possible specific effects
of the gold WEs, contacts were also grown using Pb electrodes, yielding
comparable results.

To fabricate the lead contact within the gap between the two WEs,
a potential of 10-20 mV is applied to the RE. While lead is deposited
in the junction, we monitor the conductance between the WEs. After
contact deposition, the contact is dissolved again by setting the
electrochemical potential of the RE to a value between -18 mV and
-36 mV, and closed again by choosing the potential between 6 mV and
15 mV. By continuously repeating this procedure, conductance-time
traces are recorded for a large number of opening and closing processes
at a sampling rate of 50 ms.

\begin{figure}
\begin{centering}
\includegraphics[width=0.7\columnwidth]{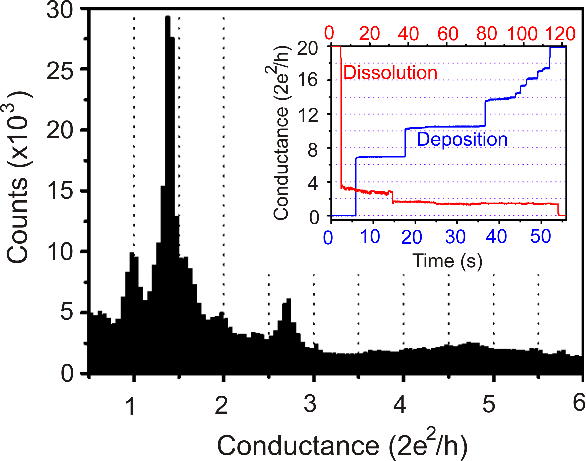}
\par\end{centering}

\caption{\label{fig:histogram}(Color online) Conductance histogram of electrochemically
fabricated atomic-sized Pb contacts. The inset shows two typical conductance-time
traces for deposition and dissolution processes. }

\end{figure}

Two typical conductance-time traces obtained for contact closing (deposition)
and opening (dissolution) are shown in the inset of Fig.~\ref{fig:histogram}.%
 The plateau regions in the traces are quite flat in general. This
fact can be explained by the lack of mechanical strain during the
electrochemical growth process, in contrast to contacts fabricated
with the MCBJ or STM techniques, where mechanical deformations are
involved \cite{Scheer:Nature1998,Cuevas:PRL1998,Pauly:PRB2006}. The
dissolution curve exhibits two plateaus at around $2.8G_{0}$ and
$1.4G_{0}$. These values are consistent with the range of the rather
continuously decreasing last conductance plateau, as reported in Refs.~\onlinecite{Scheer:Nature1998,Cuevas:PRL1998}.
The main panel of Fig.~\ref{fig:histogram} shows the conductance
histogram, plotted with a bin-size of $0.05G_{0}$. It represents
the data from $1.5\cdot10^{6}$ conductance terraces in the range
between $0$ and $20G_{0}$ for both deposition and dissolution from
64 different samples, each terrace being stable longer than 200 ms.
The structure of the histogram with its most dominant peak at $1.4G_{0}$
differs clearly from those obtained with the MCBJ approach, where
only a single, broad peak was observed between $1G_{0}$ and $3G_{0}$
and no detailed structure within this peak could be resolved \cite{Yanson:PhD2001,Makk:PRB2008}.
Here, most probably due to the lack of mechanical strain and structural
defects within the junction area, a detailed substructure is resolved
within this peak, demonstrating the advantages of the electrochemical
fabrication method. The peak position at $1.4G_{0}$, however, is
consistent with the results reported in Refs.~\onlinecite{Makk:PRB2008,Yanson:PhD2001}.
Additional peaks in the conductance distribution are found at $1G_{0}$,
$2G_{0}$, and $2.8G_{0}$. In the range from $3G_{0}$ to $6G_{0}$,
the distribution is quite flat and shows a low broad maximum centered
at around $4.8G_{0}$. 

In order to understand better the experimental findings, we have performed
a theoretical analysis of charge transport through various Pb contacts.
We describe their electronic structure at the level of DFT and determine
the conduction properties within the Landauer approach using Green\textquoteright{}s
function techniques. Employing the RI-DFT module of \noun{TURBOMOLE}
5.9 \cite{Ahlrichs:ChemPhysLett1989}, we use BP86 as exchange-correlation
functional \cite{Becke:PRA1988,Perdew:PRB1986} and the standard basis
set \textquotedblleft{}def-SVP\textquotedblright{} of split-valence
quality augmented with $d$-like polarization functions \cite{Eichkorn:TheorChemAcc1997}.
An effective core potential efficiently deals with the innermost 78
electrons \cite{Kuechle:MolPhys1991}. Our approach is described in
detail in Ref.~\onlinecite{Pauly:NJP2008}. Let us note that we determine
parameters for the description of the electrodes from a spherical
fcc cluster composed of 429 Pb atoms. The lattice constant for this
cluster is set to the experimental value of 0.495 nm. The calculation
yields $E_{F}=-3.76$ eV for the Fermi energy, a value roughly corresponding
to the experimental work function of 4.25 eV \cite{PAAnderson:PR1956}. 

\begin{figure*}
\begin{centering}
\includegraphics[width=1.4\columnwidth]{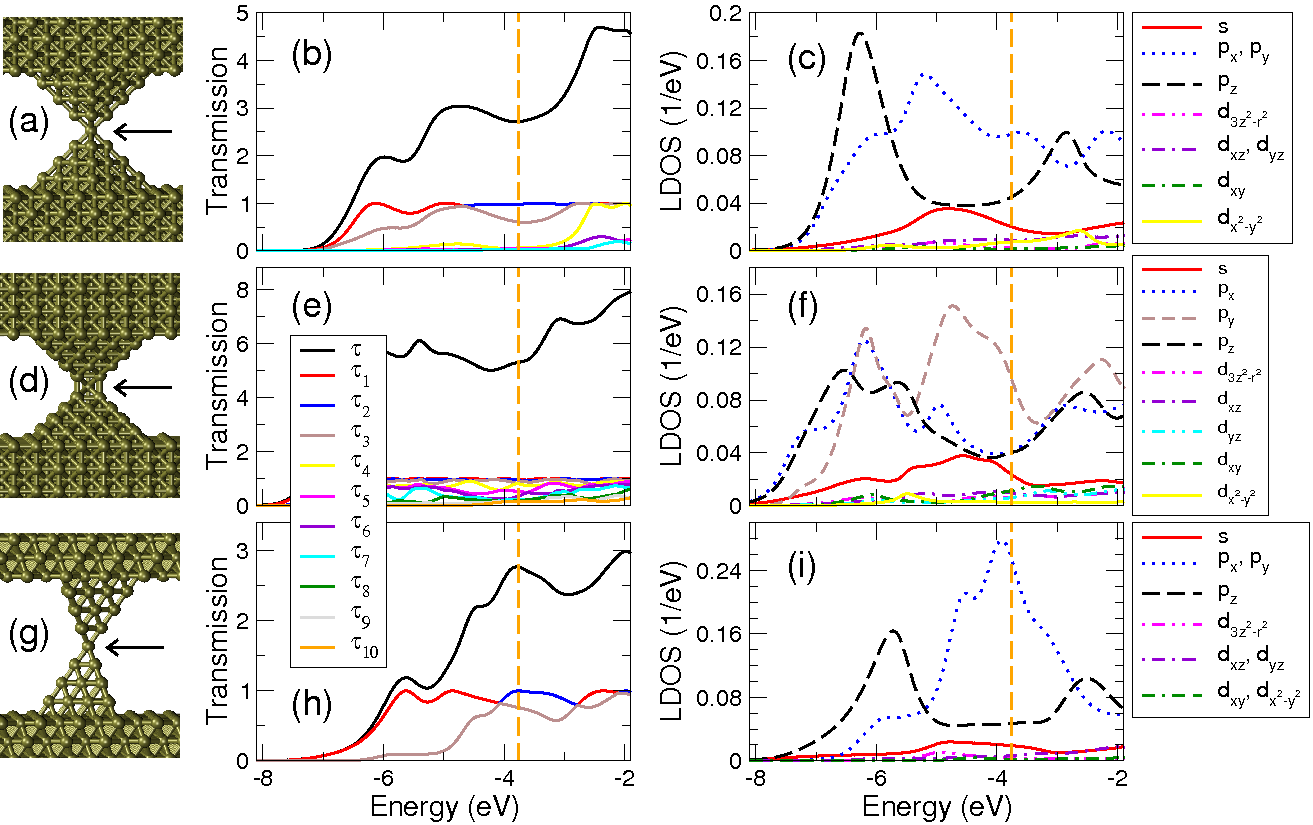}
\par\end{centering}

\caption{\label{fig:theo-single}(Color online) Geometries of single-atom contacts,
their transmission as a function of the energy, and the local density
of states (LDOS) of that atom in the narrowest part of the junction,
which is indicated by an arrow in the geometries. The transmission
$\tau=\sum_{i}\tau_{i}$ is resolved into the contributions $\tau_{i}$
from individual transmission channels, and the LDOS into its orbital
components \cite{Pauly:NJP2008}. The crystallographic orientation
of the semi-infinite electrodes is (a-c) $\left\langle 100\right\rangle $,
(d-f) $\left\langle 110\right\rangle $, and (g-i) $\left\langle 111\right\rangle $,
respectively. Vertical dashed lines indicate the Fermi energy at $E_{F}=-3.76$
eV.}

\end{figure*}

\begin{figure*}
\begin{centering}
\includegraphics[width=2\columnwidth]{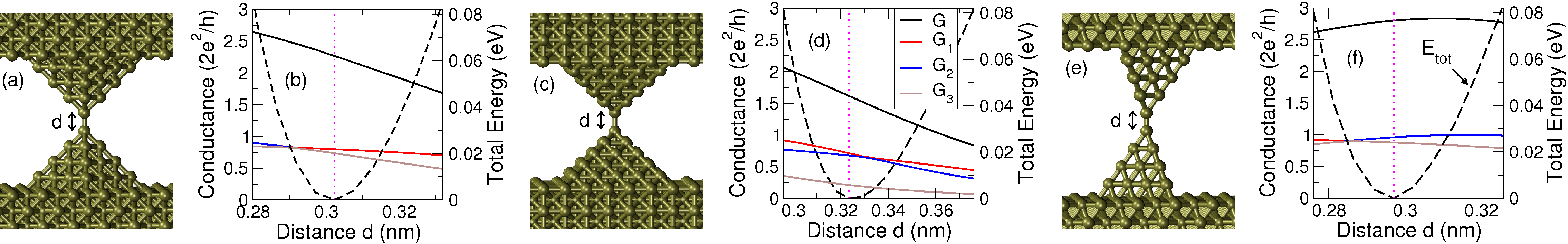}
\par\end{centering}

\caption{\label{fig:theo-dimer}(Color online) Dependence of the conductance
$G$ and the total energy $E_{tot}$ on the distance $d$ between
the tip atoms for the different dimer configurations with (a-b) $\left\langle 100\right\rangle $,
(c-d) $\left\langle 110\right\rangle $, and (e-f) $\left\langle 111\right\rangle $
crystallographic orientation. In addition to $G=\sum_{i}G_{i}$ the
contributions from the individual transmission channels $G_{i}=G_{0}\tau_{i}$
are shown. In each plot the $y$-axis on the left refers to $G$ and
$G_{i}$ (solid lines) and that on the right to $E_{tot}$ (dashed
lines). Dotted vertical lines indicate the distances $d$ with minimal
$E_{tot}$.}

\end{figure*}

We investigate two types of junction configurations, namely single-atom
and dimer contacts (Figs.~\ref{fig:theo-single} and \ref{fig:theo-dimer}).%
 For these we study different crystallographic orientations of the
electrodes, namely $\left\langle 100\right\rangle $, $\left\langle 110\right\rangle $,
and $\left\langle 111\right\rangle $. The electrodes are oriented
along the $z$ axis, the direction of charge flow. Such geometries
are believed to be responsible for the first peak in conductance histograms
of metals under ultra-high vacuum conditions. We consider only ideal
geometries, where all atoms are located in the positions of the fcc
lattice. The extended central clusters of Ref.~\onlinecite{Pauly:NJP2008},
used for the description of the contacts, are composed of around 300
atoms. To keep the numerical effort manageable, their point group
symmetry is exploited.

Results for the single-atom contacts are displayed in Fig.~\ref{fig:theo-single}.
From the transmission at the Fermi energy, we obtain conductances
of $2.7G_{0}$, $5.3G_{0}$, and $2.8G_{0}$ for the $\left\langle 100\right\rangle $,
$\left\langle 110\right\rangle $, and $\left\langle 111\right\rangle $
directions, respectively. At first glance, the conductance for the
$\left\langle 110\right\rangle $ direction is surprisingly high.
However, as is visible in Fig.~\ref{fig:theo-single}(d), this contact
should better be considered as a \textquotedblleft{}five-atom'' contact
due to the additional bonds resulting from the small distance between
atomic layers. Therefore, we will henceforth refer to single-atom
contacts as those for the $\left\langle 100\right\rangle $ and $\left\langle 111\right\rangle $
directions. For the latter two structures, the transmission at $E_{F}$
is dominated by three transmission channels. In each case, the non-degenerate
channel is of $sp_{z}$ character, while the two degenerate ones are
$p_{x}$- and $p_{y}$-like. The local density of states (LDOS) of
the atom in the narrowest part of the constriction illustrates further
the dominant role of the $6s$ and $6p$ states for conduction. 

Due to the directional character of transport the point group symmetries
relevant for the classification of transmission channels in terms
of irreducible representations are $C_{4v}$, $C_{2v}$, and $C_{3v}$
for the structures with $\left\langle 100\right\rangle $, $\left\langle 110\right\rangle $,
and $\left\langle 111\right\rangle $ orientations, respectively,
while their geometrical point groups are $D_{4h}$, $D_{2h}$, and
$D_{3d}$. Degeneracies of channels, present for the $\left\langle 100\right\rangle $
and $\left\langle 111\right\rangle $ configurations, are therefore
removed for $\left\langle 110\right\rangle $. This fact will be more
clearly visible for the dimer junctions with their lower conductance,
but the same point-group symmetries, which we shall discuss next.

In Fig.~\ref{fig:theo-dimer} we illustrate the evolution of the
conductance and of the total energy as a function of the distance
$d$ between the tip atoms for all the dimer contacts. All other distances
are kept fixed in the stretching process. At the point of minimum
total energy, conductances are $2.3G_{0}$, $1.6G_{0}$, and $2.8G_{0}$
for the $\left\langle 100\right\rangle $, $\left\langle 110\right\rangle $,
and $\left\langle 111\right\rangle $ contacts, respectively. As naively
expected, the conductance exhibits a monotonous decay with $d$ for
$\left\langle 100\right\rangle $ and $\left\langle 110\right\rangle $.
For $\left\langle 111\right\rangle $, however, it stays rather constant.
From an analysis of the LDOS of one of the chain atoms, we find that
this is due to the $p_{x}$ and $p_{y}$ states, which move into resonance.
As mentioned before, degeneracies of conduction channels are lifted
for the $\left\langle 110\right\rangle $ orientation as compared
to the other junction configurations.%

The peaks at $1.4G_{0}$ and $2.8G_{0}$ in our histogram of Fig.~\ref{fig:histogram}
are consistent with our theoretical results for single-atom and dimer
structures. This indicates that the contacts observed in the experiment
resemble ideal lead junctions without strong contaminations or structural
defects, since they would lead to deviations from these conductance
values.

The experimentally observed, pronounced peak at $1G_{0}$ is not predicted
by the conductance calculations described above. However, as the contacts
are in an aqueous electrochemical environment, small molecular species
working as bridges between the two atomic-scale Pb electrodes, such
as traces of H$_{2}$, O, O$_{2}$, or H$_{2}$O, might play a role
\cite{Smit:Nature2002,Tal:PRL2008,Makk:PRB2008}. Alternatively, ionic
lead species could be taken into account, where specific transport
channels of the junction are selectively suppressed. The interpretation
of the peak at $1G_{0}$ as a result of the conductance quantization
of a free electron gas in a geometrical constriction, as observed
in semiconductor heterostructures \cite{vanWees:PRL1988,Wharam:JPhysC:SolidStatePhysics}
and $s$-type metals \cite{Agrait:PhysRep2003}, appears as an oversimplification
for multivalent metals \cite{Yanson:PRL1997,Makk:PRB2008}.

An alternative explanation of the peaks at $2G_{0}$ and $2.8G_{0}$
would be their interpretation as replicas of those at $1G_{0}$ and
$1.4G_{0}$ with twice the value. Although geometries with parallel
atomic wires were observed for STM-fabricated gold contacts \cite{Ohnishi:Nature1998},
no indications for the formation of similar structures exist for lead
so far. 

Our analysis of transport using DFT is in agreement with theoretical
results reported in Ref.~\onlinecite{Cuevas:PRL1998}. Employing
a simpler tight-binding parametrization for describing the electronic
structure and considering only the single-atom contact oriented along
the $\left\langle 111\right\rangle $ direction, the authors report
the same conductance of $2.8G_{0}$. Keeping the central atom fixed,
they find that the conductance stays practically constant when the
junction is stretched, similar to our observations for the $\left\langle 111\right\rangle $
dimer structure. They argue that the negative slopes of the conductance
plateaus before contact rupture would require the inclusion of spin-orbit
interactions. While such relativistic effects are considered in our
calculations only in an effective manner through the core potentials
\cite{Kuechle:MolPhys1991}, our results show that the dependence
of the conductance on distance is strongly geometry-dependent.

To conclude, the high thermal and mechanical stability and low strain
in our electrochemically fabricated atomic-sized lead contacts makes
it possible to resolve sharp conductance peaks and detailed substructures
within the broad first peak in the conductance histograms obtained
with MCBJ or STM techniques. The theoretical analysis of transport
within our DFT approach yields junction geometries compatible with
the peaks at $1.4G_{0}$ and $2.8G_{0}$ and allows to attribute them
to neutral single-atom and dimer contacts of Pb. The results demonstrate
that the conductance of Pb contacts with a single-atom-wide constriction
depends crucially on the contact geometry.

F.-Q.X.\ (experiment) and F.H.\ (theory) contributed equally to
this work. We are grateful to M.~Bürkle, A.~Erbe, A.~Halbritter,
E.~Scheer, and J.\ K.\ Viljas for fruitful discussions. This work
was funded by the DFG and by the Landesstiftung Baden-Württemberg
within the Network of Excellence {}``Functional Nanostructures''.
F.P. acknowledges support through a Young Investigator Group.

\end{document}